\documentclass[preprint,pra,superscriptaddress]{revtex4-1}

\usepackage{amsmath}    
\usepackage{graphicx}   
\usepackage{color}
\usepackage[10pt]{moresize}

\usepackage{amsfonts}
\usepackage{amssymb}
\usepackage{amscd}
\usepackage{enumerate}
\usepackage{epsfig}
\usepackage{subfigure}
\usepackage{graphicx}
\usepackage{bm}

\begin{document}

\title{Experimental quantum data locking}

\author{Yang Liu}

\affiliation{Shanghai Branch, Hefei National Laboratory for Physical Sciences at Microscale and Department of Modern Physics, University of Science and Technology of China, Hefei, Anhui 230026, P.~R.~China}

\affiliation{CAS Center for Excellence and Synergetic Innovation Center in Quantum Information
and Quantum Physics, University of Science and Technology of China, Hefei, Anhui 230026, P.~R.~China}

\author{Zhu Cao}

\affiliation{Center for Quantum Information, Institute for Interdisciplinary Information Sciences, Tsinghua University, Beijing 100084, P.~R.~China}

\author{Cheng Wu}

\affiliation{Shanghai Branch, Hefei National Laboratory for Physical Sciences at Microscale and Department of Modern Physics, University of Science and Technology of China, Hefei, Anhui 230026, P.~R.~China}

\affiliation{CAS Center for Excellence and Synergetic Innovation Center in Quantum Information
and Quantum Physics, University of Science and Technology of China, Hefei, Anhui 230026, P.~R.~China}

\author{Daiji Fukuda}

\affiliation{National Metrology Institute of Japan(NMIJ),National Institute of Advanced Industrial Science and Technology(AIST), 1-1-1 Umezono, Tsukuba, Ibaraki 305-8563, Japan}

\author{Lixing You}

\affiliation{State Key Laboratory of Functional Materials for Informatics, Shanghai Institute of Microsystem and Information Technology, Chinese Academy of Sciences, Shanghai 200050, P.~R.~China}

\author{Jiaqiang Zhong}

\affiliation{Purple Mountain Observatory and Key Laboratory of Radio Astronomy, Chinese Academy of Sciences, 2 West Beijing Road, Nanjing, Jiangsu 210008, P.~R.~China}

\author{Takayuki Numata}

\affiliation{National Metrology Institute of Japan(NMIJ),National Institute of Advanced Industrial Science and Technology(AIST), 1-1-1 Umezono, Tsukuba, Ibaraki 305-8563, Japan}

\author{Sijing Chen}

\author{Weijun Zhang}

\affiliation{State Key Laboratory of Functional Materials for Informatics, Shanghai Institute of Microsystem and Information Technology, Chinese Academy of Sciences, Shanghai 200050, P.~R.~China}

\author{Sheng-Cai Shi}

\affiliation{Purple Mountain Observatory and Key Laboratory of Radio Astronomy, Chinese Academy of Sciences, 2 West Beijing Road, Nanjing, Jiangsu 210008, P.~R.~China}

\author{Chao-Yang Lu}

\affiliation{Shanghai Branch, Hefei National Laboratory for Physical Sciences at Microscale and Department of Modern Physics, University of Science and Technology of China, Hefei, Anhui 230026, P.~R.~China}

\affiliation{CAS Center for Excellence and Synergetic Innovation Center in Quantum Information
and Quantum Physics, University of Science and Technology of China, Hefei, Anhui 230026, P.~R.~China}

\author{Zhen Wang}

\affiliation{State Key Laboratory of Functional Materials for Informatics, Shanghai Institute of Microsystem and Information Technology, Chinese Academy of Sciences, Shanghai 200050, P.~R.~China}

\author{Xiongfeng Ma}

\affiliation{Center for Quantum Information, Institute for Interdisciplinary Information Sciences, Tsinghua University, Beijing 100084, P.~R.~China}

\author{Jingyun Fan}

\author{Qiang Zhang}

\author{Jian-Wei Pan}

\affiliation{Shanghai Branch, Hefei National Laboratory for Physical Sciences at Microscale and Department of Modern Physics, University of Science and Technology of China, Hefei, Anhui 230026, P.~R.~China}

\affiliation{CAS Center for Excellence and Synergetic Innovation Center in Quantum Information
and Quantum Physics, University of Science and Technology of China, Hefei, Anhui 230026, P.~R.~China}

\begin{abstract}
Classical correlation can be locked via quantum means---quantum data locking. With a short secret key, one can lock an exponentially large amount of information, in order to make it inaccessible to unauthorized users without the key. Quantum data locking presents a resource-efficient alternative to one-time pad encryption which requires a key no shorter than the message. We report experimental demonstrations of quantum data locking scheme originally proposed by DiVincenzo \emph{et al.} [Phys. Rev. Lett. \textbf{92}, 067902 (2004)] and a loss-tolerant scheme developed by Fawzi, Hayde, and Sen [J. ACM. \textbf{60}, 44 (2013)]. We observe that the unlocked amount of information is larger than the key size in both experiments, exhibiting strong violation of the incremental proportionality property of classical information theory. As an application example, we show the successful transmission of a photo over a lossy channel with quantum data (un)locking and error correction.
\end{abstract}

\maketitle

{\it Introduction.---}
Information security continuously remains the research frontier, driven by both scientific curiosity and the increasing demand from practical applications in secure communications and secure data storage. Conventionally, information security is based on computation complexity, which can be broken if equipped with enough computational capacity. Quantum mechanics fundamentally changes the game. The inherent quantum correlation enables exponential speedup in computing and unconditional information security \cite{Nielsen:2011}. Quantum key distribution \cite{Bennett1984BB84,Ekert1991QKD}, which allows two parties to generate secure keys with the help of quantum mechanics, has been demonstrated in metropolitan networks \cite{darpa03,Peev09,Sasaki11,Tang:16,maintrunkline} and is ready to commercialize. The most reliable encryption method is to encrypt the message with one-time pad \cite{otp},  where the required key size is at least as large as the size of the information. Quantum data locking allows to lock information in quantum states with exponentially shorter key, presenting an efficient solution to many resource-limited secure applications \cite{DiVincenzo04,Fawzi2013From,Lloyd:Enigma:2013, Lupo:Protocol:2014}.

The incremental proportionality of mutual information is an axiomatic property in classical information theory. Consider the following example with two parties, Alice and Bob, who start with no mutual information. First, Alice classically encodes an $n$-bit message into an $n$-bit codeword using a $k$-bit key and sends the encoded message (but not the key) to Bob. The two parties then share $n$ bit mutual information. After Alice sends the key to Bob, their mutual information increases by $k$. DiVincenzo, Horodecki, Leung, Smolin, and Terhal (DHLST)\cite{DiVincenzo04} found that a $k$-bit key can increase the mutual information by an amount more than $k$ via quantum means. This striking result of quantum data locking is due to the inherent quantum uncertainty and violates the incremental proportionality property of classical information theory in an extreme manner. Quantum data locking has received much attention since then. It was even considered to hold the potential to reconcile the black-hole information loss \cite{Leung:Survey:2009,Dupuis:Protocol:2013,Lupo:Protocol:2014}.

One of the key issues for the original quantum data locking scheme lies in the fact that message information may suffer from significant qubit loss. In 2013, Fawzi, Hayden and Sen (FHS) developed a loss-tolerant quantum data locking scheme \cite{Fawzi2013From}, in which the possible information leakage can be made arbitrarily small in a lossy environment while the unlocked information is significantly larger than the key size \cite{Guha:2014bi}. This makes quantum data locking appealing for realistic applications such as secure communication \cite{Lloyd:Enigma:2013, Lupo:Protocol:2014}.

Locking capacity is defined as the maximum accessible information to be locked with exponentially small error probability and information leakage $\epsilon$ \cite{Lupo:2015bh}. It is larger than or equal to the private capacity which is the maximum rate for secure information exchange according to the Holevo information \cite{QLKD14,Devetak:2005ge}. The main drawback for this definition is that the accessible information criterion does not assure composable security in data locking \cite{Konig:2007fv}. The composable security may be fulfilled conditional on the bounded quantum storage assumption \cite{QLKD14, Guha:2014bi} that Eve can keep his qubits only for limited time (or limited number), which is satisfied for a memoryless communication channel or the case without good quantum memories. For the later case, the two parties may perform error reconciliation after Eve's quantum memory decoheres, then the key generated by quantum locked key distribution is composably secure.

Experimental realization of quantum data locking was considered to be a technical challenge \cite{Lupo:Protocol:2014}. Here, we report experimental demonstrations of both DHLST scheme and FHS scheme with heralded single photons. We develop a robust experimental system with an overall single photon transmittance $\eta$, from preparation to detection, of $>50\%$. We employ two types of state-of-the-art superconducting single photon detectors --- superconducting nanowire single photon detector (SNSPD) \cite{Marsili:2013,Chen:15} and superconducting transition-edge-sensor (TES)  \cite{Lita08,Fukuda11} in our experiment. The fast time response of SNSPD allows encoding/decoding in real time, which is critical to the FHS scheme, and TES has high single photon detection efficiency which is necessary to fulfill the requirement to implement the DHLST scheme. The robust system allows the experiment to run continuously for over 50 hours in order to show high data locking efficiency for the FHS scheme. Besides, a comprehensive simulation with experimentally determined single photon transmittance and bit error rate ($e_{b}$) is performed to optimize the parameters of the FHS scheme. Our experimental results solidly demonstrate data locking in a variety of experimental settings, suggesting that quantum data locking has promising applications in secure communication and secure storage.
In the following, we present data locking schemes and our experimental results.

{\it Data locking schemes.---}
In the DHLST scheme, Alice encodes messages with a set of orthonormal bases and then encrypt the messages by applying a unitary operation, Identity or Hadamard transform depending on the key bit 0 or 1, to each of the qubits. In quantum information, it can be shown that the maximum amount of accessible mutual information is $n/2$ without the one-bit key; while the $n$-bit message can be completely recovered with the one-bit key.

The FHS scheme is the first explicit loss-tolerant locking scheme.  Central to this scheme is to combine  mutually unbiased bases and permutation extractors in the preparation of a set of unitaries.  The implementation of the former bounds the probability that Eve may guess the outcome of the associated measurement, and the implementation of the latter is to further distill the randomness into almost uniform bits \cite{Fawzi2013From}. A random draw from the set of unitaries is used as the key to encrypt the messages. The implementation of this scheme consists of two encoding stages. In the first stage, a block of the message is converted to the eigenstates of $Z$ (denoted as 0) or $Y$ basis (denoted as 1). The basis is set according to a Reed-Soloman code concatenated with a Hadamard code. By doing so, the Hamming distance between different messages after encryption is pair-wisely maximal. In the second stage, the produced qubit sequence is transformed with a strong permutation extractor to further optimize its difference with the original message statistically. The decoding process is a time-reversal of the encoding process. Note that the classical permutation may be performed prior to the partial Hadamard transform. (See Supplementary Material for details about the realization of FHS scheme.)

In the FHS scheme, the basis choices consumes a secret key of length $\log (2/\epsilon^2)$, and the permutation extractor consumes a key of length $40000 \log (24n^2/\epsilon)$. As shown in Supplementary Material, the mutual information is $ 6\epsilon n/16.12+H(\epsilon)$ without knowing the key, and expands to $\eta \times n/16.12 (1-H(e_{b})) $ given the secret key. Here $H(\cdot)$ is the binary Shannon entropy, and the information is calculated excluding the key.

\begin{figure*}[tbh]
\centering
\resizebox{13cm}{!}{\includegraphics{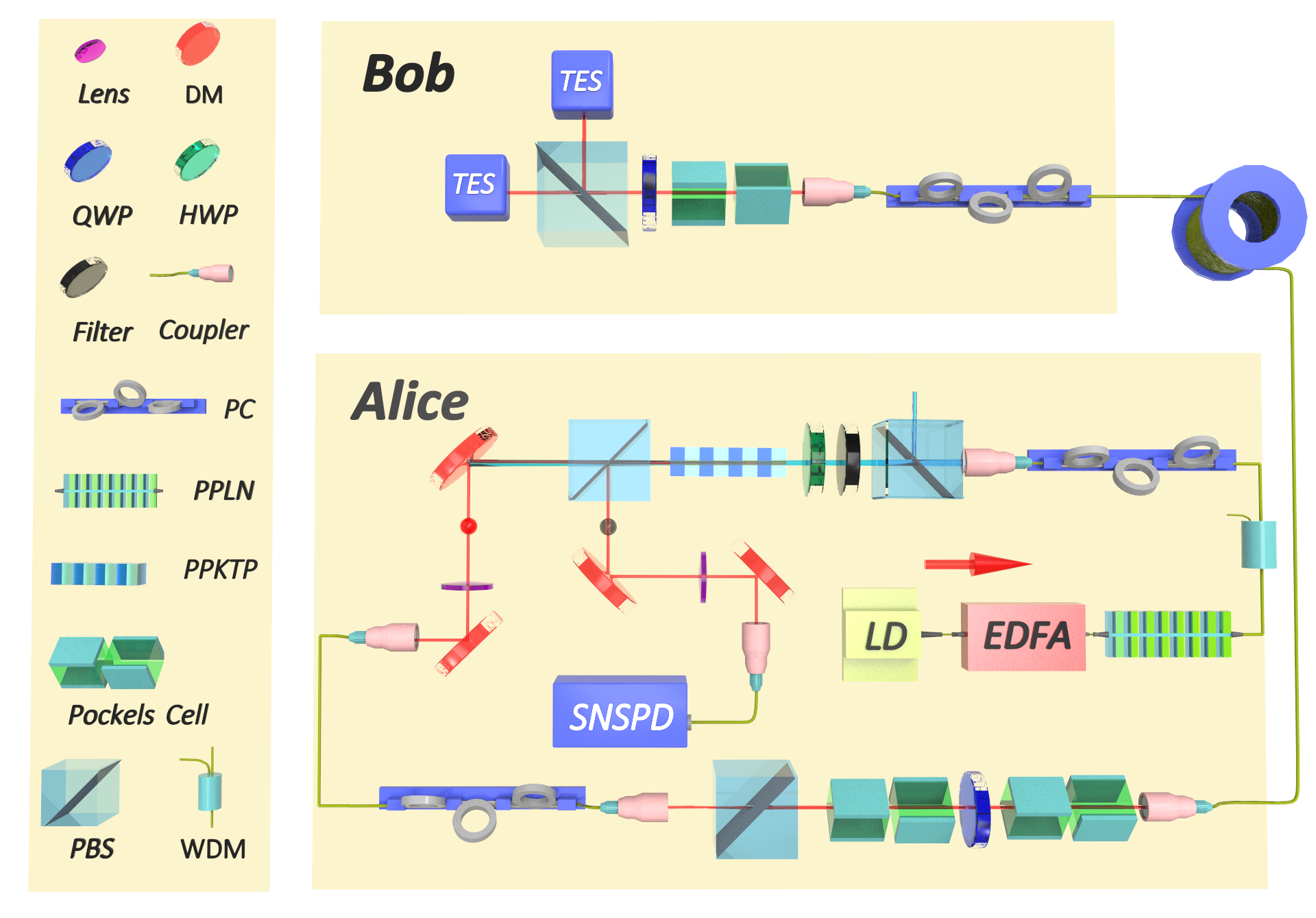}}
\caption{Schematic of experimental quantum data-locking:
Alice pulses a distributed feedback laser diode (LD) at $\lambda$=1560 nm with a pulse width of 10 ns  at 100 kHz. After passing through an erbium-doped fiber amplifier (EDFA), the laser pulses are up-converted to 780 nm via second harmonic generation (SHG) in an in-line periodically-poled lithium niobate (PPLN) waveguide.  The residual long wavelengths are removed with a wavelength-division multiplexer (WDM) and a 945 nm low pass (LP) filter. Alice focuses the pump pulses at 780 nm into a periodically-poled potassium titanyl phosphate (PPKTP) crystal to create pairs of orthogonally polarized photons that are degenerated at 1560 nm via spontaneous parametric downconversion. The photon pairs are separated by a polarizing beam splitter (PBS). Alice uses dichroic mirrors (DMs) to remove the residual pump light at 780 nm and fluorescence. The pairs of signal and idler photons are collected into single mode optical fibers. Alice detects the idler photons with a superconducting nano-wire single photon detector (SNSPD) to herald the presence of signal photons. The heralded signal photons are encoded by pockels cells. After encoding, the single photons are sent to Bob via a fiber spool. In the meantime, a control signal is sent to Bob to prepare his bases accordingly to decode the incoming single photon signals, which are received by two transition edge sensors (TES) after a PBS. Polarization controller (PC) is applied wherever it is needed to maximize transmittance of photons in the right polarization and the extinction ratio. System synchronization is controlled by a field programmable gate array (FPGA).
}
\label{Fig:Path:DataLocking}
\end{figure*}

{\it Experiment setup.---}
We experimentally demonstrate quantum data locking schemes with single photons. As shown in Fig.1, we pass single-spatial-mode 780 nm laser light through a 10 mm, periodically-poled potassium titanyl phosphate (PPKTP) crystal, which converts the pump photons into pairs of daughter photons at 1560 nm via type-II spontaneous parametric downconversion process \cite{Fedrizzi:07}. The pair of correlated, orthogonally polarized daughter photons are separated by a polarizing beam splitter (PBS) and then coupled into single mode optical fibers. We remove the residual pump photons by dichroic mirrors. We herald the presence of single photons by detecting their twin partners. With the beam waists set to be 180 $\mu$m and 85 $\mu$m for the pump and collection beams at the center of the crystal, respectively, the single photon heralding efficiency is determined to be 87\%, including all losses in the photon pair source setup \cite{Bennink10,CunhaPereira13,Dixon14} .

The experimental implementations of the two data-locking schemes are similar. Because DHLST scheme uses only 1-bit pre-shared key to choose basis $Z$ or $Y$, we use one pockels cell to encode the messages in the experiment. In FHS scheme, a time-varying bases sequence is required. We modulate the messages and bases using two successive pockels cells. As shown in Fig. 1, Alice first brings the single photons to free space and passes them through a PBS. Then she encodes the message by setting the first pockels cell to 0- or $\lambda/2$- voltage and chooses the bases by setting the second pockels cell to 0- or $\lambda/4$- voltage. Both pockels cells are initially oriented at $45^{\circ}$ with respect to the vertical axis. When applied with $\lambda/2$- or $\lambda/4$- voltage, the first pockels cell functions as a half-wave plate  oriented at $45^{\circ}$ and the second pockels cell functions as a quarter-wave plate oriented at $45^{\circ}$. After encoding, the photons are coupled into single mode fibers for delivery.  Bob uses a pockels cell to set his bases similarly by applying 0- or $\lambda/4$- voltage. The loss in the encoding (decoding) process is determined to be 7\%, which is mainly due to the mismatch between free space optical mode and fiber optical mode.

We use a SNSPD with a rising edge of $\tau\sim$70 ps as the heralding detector. The fast timing response allows to orient pockels cells appropriately to encode/decode messages in real time, and the relatively high detection efficiency (50\%) helps to create a good rate of single photons to reduce the running time of experiment. We use a TES to detect the signal photons at the receiver. The single photon detection efficiency of TES is determined to be 75\% when held at $\sim$100 mK.

A field-programmable-gate-array (FPGA) provides a 100 kHz signal to pulse the pump laser. Upon receiving the heralding signals, the FPGA sends signals to pockels cells (to prepare bases and unitary operations) to encode the heralded single photons with quantum states according to the pre-programmed data locking scheme. The FPGA also sends signals to prepare Bob's pockels cell to decrypt the message according to the pre-shared key, such that the received single photons are detected in the correct bases by TES.

We note that the permutation step is a classical algorithm and does not affect the performance of the data locking schemes. The fully realization is left to the future work. We have nevertheless taken into account the seed consumption of this permutation step in the data analysis.

{\it Experimental results.---}
We first realize the DHLST scheme. We set the basis to be $Z$($Y$) if the key is 0(1), and send more than 8 Mb data in each basis. As shown in Table \ref{tab:divin}, for both bases, single photon transmittance, from preparation in Alice's station to detection in Bob's station is determined to be greater than 55\%, the measured error rate is less than 0.4\%. The accessible mutual information ($I_{acc}(A:B)$ ) between Alice and Bob is greater than the maximum amount of information ($n/2$) that a receiver who does not have the key, which clearly exhibits data locking.

\begin{table}[htb]
\centering
  \caption{Experimental results of data locking with DHLST scheme ($\sigma$ represents 1-standard deviation).}
\begin{tabular}{|c|c|c|c|}
\hline
 &  $e_{b}$  & $\eta$ & $I_{acc}(A:B)/n \pm \sigma$ \\
\hline
$Z$ basis & $0.4\%$ & $55.2\%$ &   $53.1\pm 0.4 \%$ \\
$Y$ basis & $0.3\%$ & $56.6\%$  &  $54.9 \pm  1.4 \%$ \\
\hline
\end{tabular}
\label{tab:divin}
\end{table}

To experimentally demonstrate the loss-tolerant FHS scheme, the single photon transmittance is tailored to be $54\%$, $41\%$ and $33\%$ by setting the fiber length accordingly to be 0 km, 5 km and 11 km. For each length, we vary the data size from $64$ Mb to $640$ Mb to examine the data locking. By setting $\epsilon=10^{-9}$, Eve's accessible information $I_{acc}(A:E)$ is bounded by 1; while $I_{acc}(A:B)$ is proportional to $n$ (See Supplementary Material for detail).

\begin{figure*}[htb]
\centering
    \subfigure[]{
      \includegraphics[width=6cm]{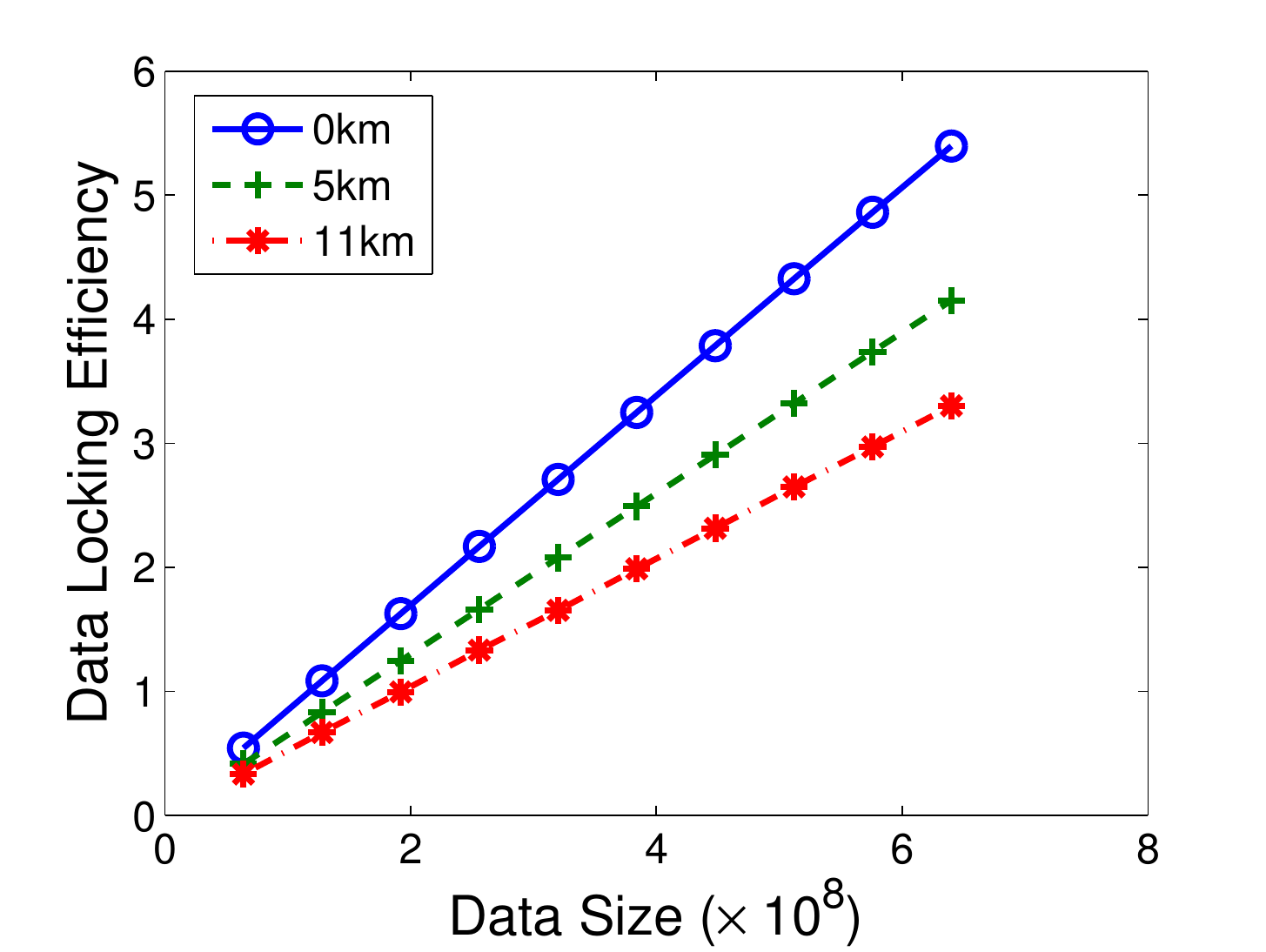}
    }
    \subfigure[]{
      \includegraphics[width=5cm]{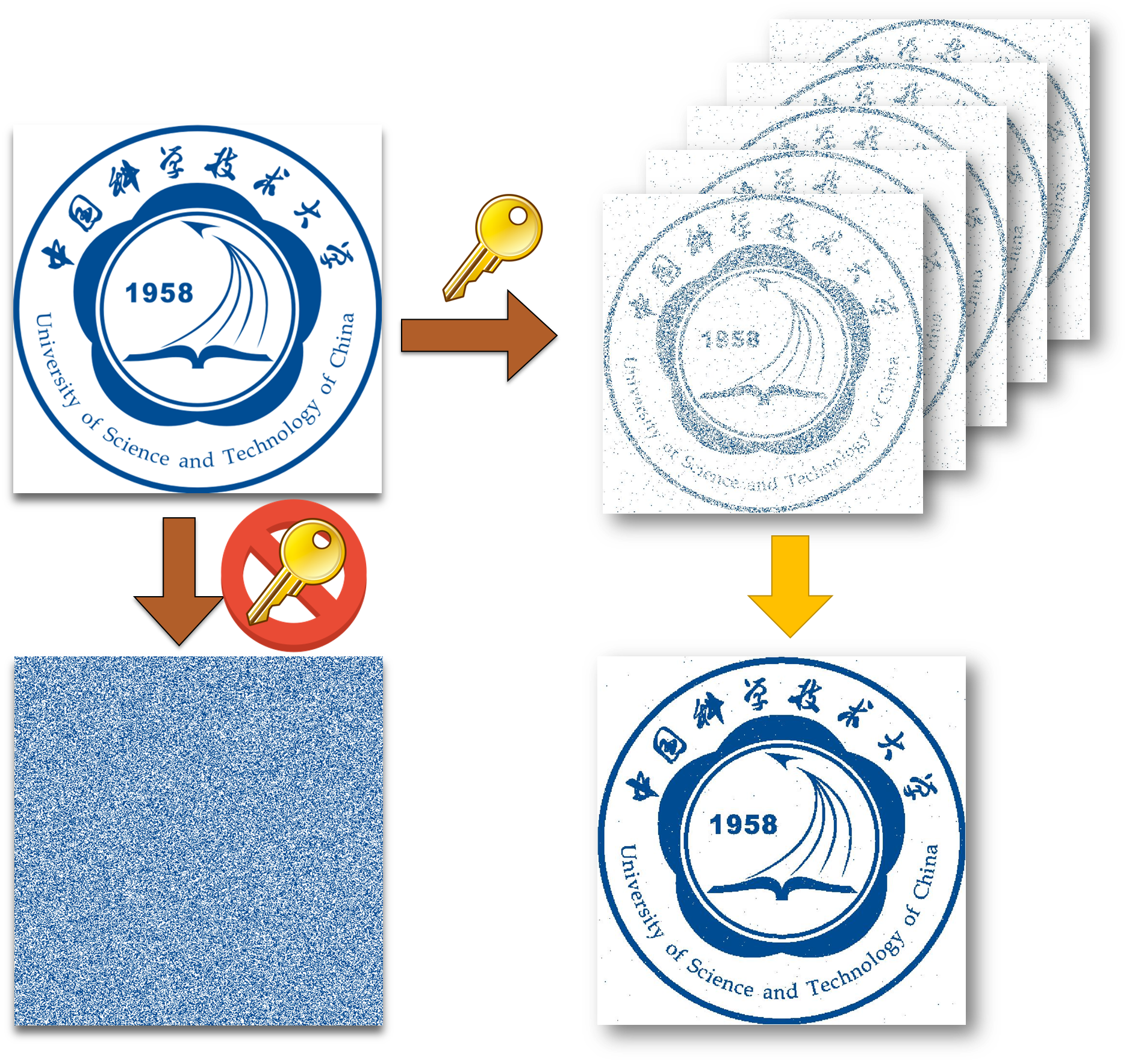}
    }
    \subfigure[]{
      \includegraphics[width=6cm]{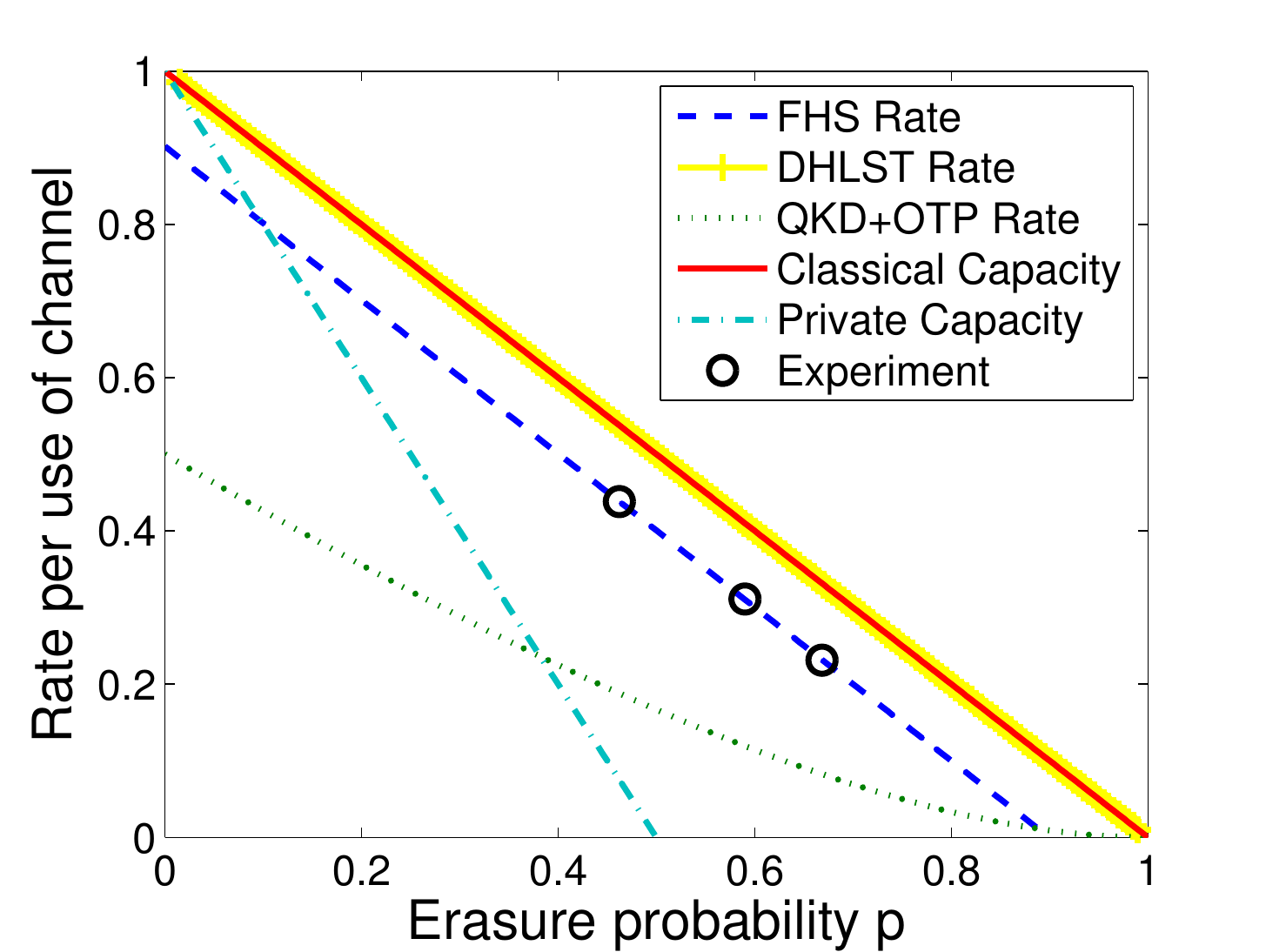}
    }
\caption{(color online)
(a) Data locking efficiency of FHS scheme with tailored single photon transmittance.
(b) Sending a photo with data (un)locking and error correction.
(c) Communication rate in a quantum erasure channel.  }
\label{fig:capacity}
\end{figure*}

We define the data locking efficiency as
\begin{equation}
\kappa= \frac{I_{acc}(A:B)-I_{acc}(A:E)-r}{r},
\end{equation}
where $r$ is the key length,  $I_{acc}(A:E)$ and $I_{acc}(A:B)$ are the mutual information before and after reconciliation between Alice and Bob.

The data locking efficiency grows linearly with data size, as shown in Fig.~\ref{fig:capacity}(a). It requires larger data size to surpass the performance of one time pad (with $\kappa = 1$) as the system loss increases. For our experiment, the data locking efficiency beats the performance of one time pad when data size is larger than 128 Mb, 192 Mb and 256 Mb for $\eta=$ $54\%$, $41\%$, and $33\%$, respectively.

Information integrity is also critical in secure applications. Here we  realize forward error correction (FEC) with erasure coding in the experimental implementation of quantum data locking. As an example, we send a photo of the logo of University of Science and Technology of China with quantum data (un)locking through a lossy channel. We repeat each encoded qubit by $50/\eta$ times. As such, we can recover each qubit with a probability of $1-(1-\eta)^{50/\eta}\ge 1-exp(-50)$; while Eve's information increases only by $50/\eta$ times. As shown in Fig.~\ref{fig:capacity}(b), with the key, the photo of the logo at the receiver is sharp with error correction code as compared to the blurred one without using error correction code. Without the key, what is received is simply a set of random data.

An important application of data locking is quantum-locked key distribution. We estimate the performance of key-distribution based on our experimental results (open circle) with $\epsilon = 10^{-9}$ , and compare it with classical capacity and private capacity. Here the classical capacity is the maximum amount of information that can be sent through the channel regardless of security. The private capacity is the secure part of the information when sending the information directly through the channel without any encoding. For a qubit erasure channel, the private capacity is $1-2p$ and the classical capacity is $1-p$, where $p$ is the erasure probability. As shown in Fig.~\ref{fig:capacity}(c), the secure communication rate of data locking (long dashed line) is well above the private capacity (dotted-dashed line) and is close to the classical capacity (solid line). We also plot the estimated secure key rate based on DHLST scheme (thick solid line) in Fig.~\ref{fig:capacity}(c), which basically overlaps with the classical capacity by consuming only one additional bit. For comparison, we plot the secure key rate of the most-used QKD+one time pad (OTP) combination (dashed line, see Supplementary Material), which is less than one half of the rate based on data locking. The difference will be even larger when transmitting a longer random number sequence using quantum locked key distribution. However, we note that in terms of security,  QKD+OTP is better than quantum locked key distribution using the FHS scheme (which is much higher than using DHLST scheme). Yet, the security of quantum locked key distribution using the FHS scheme with bounded quantum storage assumption can be as good as QKD.

{\it Conclusion.---}
In conclusion, we have experimentally shown  data locking with single photons in a variety of experimental settings. Our analysis shows its potential in key distribution. As an example for future applications, we successfully transmitted  a photo with data (un)locking and error correction code. Our experimental results suggest that quantum data locking holds potentials in many resource-limited secure information applications.

The authors would like to thank J.-Y.~Guan, L.-K.~Chen, Y.-H.~Li and Q.-C.~Sun for enlightening discussions.
This work has been supported by the National Fundamental Research Program (under Grants No. 2011CB921300 and No. 2013CB336800), the National Natural Science Foundation of China, the Chinese Academy of Science, and the 1000 Youth Fellowship program in China.

\bibliographystyle{apsrev4-1}
\bibliography{Biblilock}

\end{document}